%Paper: hep-th/9309127
%From: dowker@v2.ph.man.ac.uk
%Date: Thu, 23 Sep 1993 10:37:01 +0100
%Date (revised): Thu, 23 Sep 1993 11:46:25 +0100
%Date (revised): Thu, 23 Sep 1993 12:03:42 +0100

\magnification=1200
\hsize=6truein\vsize=8.5truein
 %scaled\magstep1 % For VAX
%\font\open=msym10 %scaled\magstep1 % For Arbortxt on PC
  % For Arbortxt on PC and VAX

\font\bigbf=cmbx10 scaled\magstep1

\def\mbox#1{{\leavevmode\hbox{#1}}}

\def\la{\lambda}
\def\si{\sigma}

     % Newline

\def\frac#1/#2{\leavevmode\kern.1em
\raise.5ex\hbox{\the\scriptfont0 #1}\kern-.1em/\kern-.15em
\lower.25ex\hbox{\the\scriptfont0 #2}}
\def\sfrac#1/#2{\leavevmode\kern.1em
\raise.5ex\hbox{\the\scriptscriptfont0 #1}\kern-.1em/\kern-.15em
\lower.25ex\hbox{\the\scriptscriptfont0 #2}}

\def\gtorder{\mathrel{\raise.3ex\hbox{$>$}\mkern-14mu
             \lower0.6ex\hbox{$\sim$}}}
\def\ltorder{\mathrel{\raise.3ex\hbox{$<$}|mkern-14mu
             \lower0.6ex\hbox{\sim$}}}

\def\semidirprod{\rlap{\ss C}\raise1pt\hbox{$\mkern.75mu\times$}}

\def\for{\lower6pt\hbox{$\Big|$}}
\def\fish{\kern-.25em{\phantom{abcde}\over \phantom{abcde}}\kern-.25em}

\def\boxit#1{\vbox{\hrule\hbox{\vrule\kern3pt
        \vbox{\kern3pt#1\kern3pt}\kern3pt\vrule}\hrule}}
\def\dalemb#1#2#3{{\lower#3pt\vbox{\hrule height .#2pt
        \hbox{\vrule width.#2pt height#1pt \kern#1pt
                \vrule width.#2pt}
        \hrule height.#2pt}}}
\def\Sbox{\mathord{\dalemb{5.9}{6}{0.5}\hbox{\hskip1pt}}}
\def\square{\mathord{\dalemb{5.9}{6}{1}\hbox{\hskip1pt}}}

        %double stroke
 %lower covariant deriv.
    %lower ordinary  deriv.

\def\noin{\noindent}

\def\de{\delta}

\def\la{\lambda}

\def\om{\omega}
\def\si{\sigma}

\def\ze{\zeta}

\def\ie{{\it i.e. }}
\def\cf{{\it cf }}

   %gives bracket <#1|#2>
  %gives matrix element <#1|#2|#3>

\def\gap{\vskip 20truept}

\def\sect{{\vskip 10truept\noindent}}

\def\3j#1#2#3#4#5#6{\left\lgroup\matrix{#1&#2&#3\cr#4&#5&#6\cr}
\right\rgroup}

%-------------------- Equation numbering macro

\def\nolabels{\def\eqnlabel##1{}\def\eqlabel##1{}\def\reflabel##1{}}
\def\writelabels{\def\eqnlabel##1{%
{\escapechar=` \hfill\rlap{\hskip.09in\string##1}}}%
\def\eqlabel##1{{\escapechar=` \rlap{\hskip.09in\string##1}}}%
\def\reflabel##1{\noexpand\llap{\string\string\string##1\hskip.31in}}}
\nolabels
\global\newcount\meqno \global\meqno=1
\global\meqno=1
\def\eqnn#1{\xdef #1{(\the\meqno)}%
\global\advance\meqno by1\eqnlabel#1}
\def\eqna#1{\xdef #1##1{\hbox{$(\the\meqno##1)$}}%
\global\advance\meqno by1\eqnlabel{#1$\{\}$}}
\def\eqn#1#2{\xdef #1{(\the\meqno)}\global\advance\meqno by1%
$$#2\eqno#1\eqlabel#1$$}
%% Style is \eqn\label{Equation}.

%--------------------- Reference macros:

%---\ref\label{text} generates number to \label, generates an entry.
%---\refn\label{text} = \ref\label{text} without written number on page.
%---\listrefs: list the refs on a separate page,
\global\newcount\refno
\global\refno=1 \newwrite\reffile
\newwrite\refmac
\newlinechar=`\^^J
\def\ref#1#2{\the\refno\nref#1{#2}}
\def\nref#1#2{\xdef#1{{\bf\the\refno}} %I have made the text ref.numbers bold.
\ifnum\refno=1\immediate\openout\reffile=refs.tmp\fi
\immediate\write\reffile{
     \noexpand\item{[{\noexpand#1}]\ }#2\noexpand\nobreak.}
     \immediate\write\refmac{\def\noexpand#1{\the\refno}}
   \global\advance\refno by1}
\def\semi{;\hfil\noexpand\break ^^J}
\def\refn#1#2{\nref#1{#2}}

\def\pl#1#2#3{{\it Phys. Lett.} {\bf {#1}B} (19{#2}) #3}
\def\np#1#2#3{{\it Nucl. Phys.} {\bf B{#1}} (19{#2}) #3}

%  *****************     Reference list     *************************
\refn\Duff{M.Duff {\it Twenty years of the Weyl anomaly} CTP-TAMU-06/93,
(1993)}
\refn\Luscher{M.L\"uscher, K.Symanzik and P.Weiss \np {173}{80}{365}}
\refn\Polyakov{A.Polyakov \pl {B 103}{81}{207}}
\refn\Bukhb{L.Bukhbinder, V.P.Gusynin and P.I.Fomin {\it Sov. J. Nucl.
 Phys.} {\bf 44} (1986) 534}
%reffs
% ********************************************************************

\vglue 1truein
\rightline {MUTP/93/22}
\gap
\centerline {\bigbf A note on Polyakov's nonlocal form}
\centerline {\bigbf of the effective action}
\vskip 15truept
\centerline{J.S.Dowker}
\vskip 10 truept
\centerline {\it Department of Theoretical Physics,}
\centerline{\it The University of Manchester, Manchester, England.}
\vskip 40truept
\centerline {Abstract}
\vskip 10truept
A technical point regarding the invariance of Polyakov's nonlocal form
of the effective action under uniform rescalings is addressed.
\vfill\eject
\noin{\bf 1. Introduction}.

\noin The following is a brief note on the conformal anomaly and its
integration
to give the change in the effective action under a conformal transformation.
This is a well--trodden path, especially in two dimensions, and no attempt
will be made here to give proper references. A recent review, with a
personal
flavour, by Duff, [\Duff], provides a reasonable, historical perspective
on the conformal anomaly.

In two dimensions, which is what concerns us here, the anomaly was
integrated to give the effective action by L\"uscher, Symanzik and Weiss
[\Luscher], and by Polyakov [\Polyakov], long ago. The point at issue
is Polyakov's conversion of the local form of the effective action to a
non-local expression when there is a zero mode, as for a closed space.
We can therefore ignore boundary effects.

The conclusions of this paper are elementary and probably
known to workers in the field. However, the author has not been able
to find a suitable published discussion.
\sect{\bf 2. Integrating the anomaly}

\noin The conformal scaling is $g_{\mu\nu}\rightarrow\bar g_{\mu\nu}=
\la^2 g_{\mu\nu}=\exp(-2\om)g_{\mu\nu}$ under an infinitesimal example of
which the renormalised effective action, $W$, changes by
\eqn\basic{
\de W[\bar g]=-\ze\big(0,\bar g;\de\om\big)
}
with
\eqn\fold{
\ze\big(0,\bar g;f\big)\equiv\int_{\overline{\cal M}}\ze\big(0,\bar g,x\big)
f(x)\,(\bar g)^{1/2}\,dx.
}

Standard theory gives the local value
\eqn\zetazero{
\ze\big(0,g,x\big)={1\over24\pi}R-P_0
}
where $P_0(x)$ is the projection onto the zero mode. In the following we
will set
$P_0=1/A$ where $A$ is the area of the closed 2-manifold ${\cal M}$ of metric
$g$. The manifold of metric $\bar g$ is denoted by ${\overline{\cal M}}$.

The idea is to integrate \basic\ knowing the explicit dependence of the
right-hand side on the scaling function $\om$. As already said, the
result is very old. It is

\eqn\effact{
W[\bar g,g]=-{1\over2}\ln\big({A[\bar g]\over A[g]}\big)
-{1\over24\pi}\int\om(R+\square\om)
g^{1/2}\,dx
}
where $\Sbox$ is the covariant Dalembertian.

In order to get a symmetrical formula in terms of the geometry corresponding
to the two metrics, we can try to eliminate the conformal factor $\om$
by solving the conformal relation
\eqn\reln{
\square\om={1\over2g^{1/2}}\big(\bar g^{1/2}{\bar R}-g^{1/2}R\big)
}
for $\om$. This is possible because the right-hand side is orthogonal in
${\cal M}$ to the zero mode, a uniform function, by topological invariance. The
solution then reads
\eqn\soln{
\om(x)={1\over2}\int G(x,y)\big(\bar g^{1/2}{\bar R}-g^{1/2}R\big)_y\,dy
+{1\over A}\int\om(y)g^{1/2}\,dy,
}
where the Green function $G$ is to be thought of as having the zero mode
removed, \ie it satisfies the standard equation
\eqn\gf{
\square G(x,y)=\de(x,y)-{1\over A}.
}
Note that removing the zero mode destroys the equality of $G$ and $\bar G$,
except under uniform rescalings. $\bar G$ obeys
\eqn\gfb{
{\overline{\square}} \bar{G}(x,y)={\bar\de(x,y)}-{1\over\bar A},
}
with
\eqn\dalinv{
\bar g^{1/2}\,{\overline{\square}}=g^{1/2}\,\square.
}

It should be pointed out that $\de(x,y)$ is the covariant delta
function on ${\cal M}$. It equals $\de_P(x,y)/g^{1/2}$, where $\de_P(x,y)$ is
the metric-independent, Dirac delta function periodised appropriately for a
closed manifold.

Substituting for $\om$ in \effact, and using \reln, we obtain (\cf [\Bukhb]
but without the zero mode contribution)
$$
W[\bar g,g]=-{1\over2}\ln\big({A[\bar g]\over A[g]}\big)
-{1\over48\pi A}\int\ln\big({g\over\bar g}\big)\, g^{1/2}\,dx\int\!
Rg^{1/2}dx$$
\eqn\effactb{
-{1\over96\pi}\int\big(\bar g^{1/2}{\bar R}+g^{1/2}R\big)_xG(x,y)\,
\big(\bar g^{1/2}{\bar R}-g^{1/2}R\big)_y\, dx\,dy.
}

The first term on the right-hand side of \effactb\ is the standard zero
mode contribution. The second term is our (possible)
novelty which goes part way to relieving the unease expressed
by Duff [\Duff] regarding the variation (or rather, non--variation) of the
nonlocal action under uniform rescalings. Under such, the final term in
\effactb\ is unchanged but the second yields the required variation.

The second term on the right-hand side of \effactb\ (and of its symmetrical
form) can be removed if $\om$ is constrained to being orthogonal to the
zero modes in both ${\cal M}$ and ${\overline{\cal M}}$. Of course, this then
eliminates the possibility of discussing uniform rescalings in the nonlocal
formulation.

It is not obvious that one can write the
right-hand side of \effactb\ as $W[\bar g]-W[g]$. But, using \dalinv\ and
solving \reln\ in terms of $\bar G$, we find an equivalent form for $W$ that
implies the required antisymmetry, $W[\bar g,g]=-W[g,\bar g]$. The second
cocycle condition, $W[g_1,g_2]+W[g_2,g_3]+W[g_3,g_1]=0$, is also true.
This implies the existence of the functional $W[g]$. It is probably easiest to
show this from the local expression \effact\ by noting that
\eqn\dblderiv{
{\de^2\over\de\om\de\si}W\big[e^{-2\om}g,e^{-2\si}g\big]=0.
}

Unlike Polyakov's case, it does seem possible to disentangle $\bar g$ from $g$
in \effactb\ but, formally, we could define $W[g]$, up to an additive
constant, by $W[g]=W[g,g_0]$ where $g_0$ is some fiducial metric.

\sect{\bf 3. Conclusion}

\noin Our unexceptional conclusion is that one must be careful when
employing Polyakov's nonlocal form of the effective action,
$W[g]\approx\int R\,\Sbox^{-1}R$. A too cavalier introduction could
fabricate difficulties such as those concerning uniform scalings.
\vskip 30truept

%  \vfill\eject\immediate\closeout\reffile%\parindent=20pt
  \immediate\closeout\reffile%\parindent=20pt % no eject
%  \centerline{{\bf References}}\bigskip\frenchspacing%  Central heading
  \noindent{{\bf References}}\bigskip\frenchspacing% Left heading

  \input refs.tmp\vfill\eject\nonfrenchspacing

\end